\documentclass[aps,pre,reprint,superscriptaddress,amsmath,amssymb,showpacs,floatfix,longbibliography]{revtex4-1}
\usepackage{graphicx}
\usepackage{bm}
\usepackage{color}
\usepackage{lipsum}

\begin{document}

\title{Universal Property of the Housekeeping Entropy Production}

\author{Hyun-Myung Chun}
\affiliation{II. Institut f{\"u}r Theoretische Physik, Universit{\"a}t Stuttgart,
70550 Stuttgart, Germany}
\author{Jae Dong Noh}
\affiliation{Department of Physics, University of Seoul, 02504 Seoul,
Korea}

\date{\today}

\begin{abstract}
The entropy production of a nonequilibrium system with broken
detailed balance is a random variable whose mean value is
nonnegative. Among the total entropy production, the housekeeping entropy 
production is associated with the heat dissipation in maintaining a
nonequilibrium steady state. 
We derive a Langevin-type stochastic differential equation for the 
housekeeping entropy production. 
The equation allows us to define a housekeeping entropic time $\tau$.
Remarkably, it turns out that the probability distribution of the 
housekeeping entropy production at a fixed value of $\tau$ is given by the 
Gaussian distribution regardless of system details. The Gaussian
distribution is universal for any systems, whether in the steady state or in
the transient state, whether they are driven by
time-independent or time-dependent driving forces.
We demonstrate the universal distribution numerically for model systems.

\end{abstract}
\maketitle

\section{Introduction}\label{sec:introduction}

Recent developments of stochastic thermodynamics have made it possible to
investigate thermodynamic properties of small-sized nonequilibrium 
systems~\cite{Seifert:2012es}.
In stochastic thermodynamics, thermodynamic quantities are defined
at the level of a single trajectory realized by stochastic dynamics~\cite{Sekimoto:1998uf}.
Among them, the entropy production is of particular importance.
The total entropy production consists of two contributions: change in the 
stochastic entropy of a system under consideration and the entropy production 
in the thermal environment surrounding the system~\cite{Seifert:2005vb}.
The entropy production turns out to be a measure of the extent 
to which the time-reversal
symmetry is broken~\cite{Schnakenberg:1976wb,Lebowitz:1999tv,Maes:2003tc,Chun:2018bk}.

The total entropy production is decomposed into two parts.
A nonequilibrium steady state is accompanied by a heat dissipation 
resulting in an entropy production of the thermal environment.
Such an entropy production to maintain a nonequilibrium steady state 
is called the housekeeping entropy production.
The remaining part is called the excess 
entropy production~\cite{Oono:1998uj,Hatano:2001uc,Lee:2013to}.
While the housekeeping entropy production is associated with dissipation 
in a nonequilibrium steady state, the excess entropy production occurs due
to extra dissipation during relaxation toward a steady state or
transition between steady states.
They are also interpreted as adiabatic and nonadiabatic parts, respectively, 
in the perspective of time-scale separation~\cite{Esposito:2010bd}.

The total entropy production exhibits intriguing universal properties.
Recently, it was found that the probability distribution should satisfy the 
fluctuation theorem. It is a direct consequence of the relation between 
the total entropy production and the broken time-reversal symmetry of 
nonequilibrium system
dynamics~\cite{Lebowitz:1999tv,Crooks:1999ta,Seifert:2005vb}. The
fluctuation theorem guarantees that the mean total entropy production should
be nonnegative. 
Experimental results support the validity of the fluctuation theorem for various 
kinds of nonequilibrium
systems~\cite{Wang:2002hw,Carberry:2004fk,Wang:2005fe,Wong:2017ge}.
More recently, after a pioneering investigation of Barato and
Seifert~\cite{Barato:2015kq}, it was found that the thermodynamic currents
should obey the universal inequality called the thermodynamic uncertainty
relation~\cite{Gingrich:2016ip,Gingrich:2017jm,Pietzonka:2017iq,Horowitz:2017ut,
Proesmans:2017ic,Dechant:2018ga,Dechant:2018vu}.
The inequality reveals a trade-off relation between the current fluctuation
and the total entropy production in a nonequilibrium steady state.
Besides, extreme-value statistics of the total entropy
production~\cite{Neri:2017vs} and entropic bounds on currents based on the Cauchy-Schwarz inequality~\cite{Dechant:2018wp} have been studied.

One of the most interesting recent investigations on the total entropy production
in nonequilibrium steady state was performed
by Pigolotti et al.~\cite{Pigolotti:2017fs}.
They derived a Langevin-type stochastic differential equation of the total entropy 
production.
In the steady state, by introducing a so-called entropic time, the stochastic
differential equation can be transformed into a universal form that does not depend 
on system details.
As a consequence, the probability distribution of the total entropy production 
at a fixed entropic time is universally given by the Gaussian distribution.
The mapping to the entropic time provides an elegant explanation 
for the origin of the universal properties of the total entropy production.
However, it is limited to the steady state of nonequilibrium 
systems under time-independent driving forces~\cite{Pigolotti:2017fs}. 
It raises a question whether the universal property can be found even in the
transient state under the time-dependent driving forces.
In this paper, we will show that the housekeeping entropy production has 
the universal property not only in the steady state but also in the transient
state even under time-dependent driving forces.

This paper is organized as follows.
In Sec.~\ref{sec:Theory}, we introduce a broad class of nonequilibrium
thermodynamic systems described by the overdamped Langevin equation, and 
derive a stochastic differential equation for the housekeeping entropy 
production.
Using the stochastic differential equation, we show that the housekeeping 
entropy production follows the universal distribution in Sec.~\ref{sec:Shk}.
In Sec.~\ref{sec:Examples}, we present numerical results for the model
systems to demonstrate the theoretical results.
We summarize our results in Sec.~\ref{sec:Conclusion}.

\section{Stochastic differential equation for the housekeeping entropy production}\label{sec:Theory}
We consider a multi-dimensional overdamped Langevin system 
in thermal contact with a heat bath at temperature $T$. 
The configuration is described by the column vector 
$\bm{x} = (x_1,\cdots,x_d)^{\rm T}$ for position variable
with the superscript $^{\rm T}$ representing the transpose.
The position variable may stand for the coordinates of a single particle in
a $d$ dimensional space or those of $n$ particles in a $(d/n)$ dimensional
space. The particle is applied to the force 
\begin{equation}\label{total_force}
\bm{F}(\bm{x},\bm{\lambda}(t),\bm{\kappa}(t)) = -\bm{\nabla}_{\bm x}
V(\bm{x},\bm{\lambda}(t)) + \bm{f}(\bm{x},\bm{\kappa}(t)) , 
\end{equation}
where the first term is a conservative force with a potential energy
$V$ and the second term is a nonconservative force. 
Note that 
$\bm{\nabla}_{\bm x} = (\partial_{x_1},\cdots,\partial_{x_d})^{\rm T}$ 
denotes the gradient with respect to $\bm{x}$ 
and that the column vector notation is adopted for $\bm{F}$ and $\bm{f}$.
In general, both forces
may depend explicitly on time through sets of protocol parameters $\bm{\lambda}(t)
\equiv \{\lambda_1(t),\cdots,\lambda_p(t)\}$ and
$\bm{\kappa}(t) \equiv \{\kappa_1(t),\cdots,\kappa_q(t)\}$.
The Langevin equation is given by
\begin{equation}\label{eq:Langevin}
\dot{\bm{x}}(t) = \mathsf{M} \bm{F}(\bm{x}(t),\bm{\lambda}(t),\bm{\kappa}(t)) 
+ \bm{\xi}(t)
\end{equation}
where $\mathsf{M}$ is the mobility matrix, 
$\bm{\xi}(t) = (\xi_1(t),\cdots,\xi_d(t))^{\rm T}$ is a Gaussian white noise
satisfying $\langle \bm{\xi}(t)\rangle = 0$ and 
$\langle \bm{\xi}(t)\bm{\xi}^{\rm T}(t')\rangle = 2\mathsf{D}\delta(t-t')$
with a diffusion matrix $\mathsf{D}$. 
The mobility matrix and the diffusion matrix are symmetric and 
positive definite, and satisfy the Einstein relation 
$\mathsf{D} = T\mathsf{M}$~\cite{Risken:1996vl,Zwanzig:2001vd}.  
They are assumed to be independent of $\bm{x}$ and $t$.
We set the Boltzmann constant to unity throughout this paper.
The Fokker-Planck equation corresponding to \eqref{eq:Langevin} is given 
by~\cite{Risken:1996vl,Gardiner:2010tp}
\begin{equation}\label{eq:FK_equation}
\partial_t P(\bm{x},t;\bm{\lambda}(t),{\bm \kappa}(t))
= -\bm{\nabla}_{\bm x} \cdot \bm{J}(\bm{x},t;\bm{\lambda}(t),{\bm \kappa}(t))
\end{equation}
with the probability current
\begin{equation}\label{eq:def_current}
\bm{J} = \mathsf{M} \bm{F}P - \mathsf{D} \bm{\nabla}_{\bm x}  P.
\end{equation}
For the sake of brevity, we omit arguments of functions unless it causes 
confusion.

The total energy $E$ of an overdamped Langevin system is given solely by the
potential energy. When the protocol $\bm{\lambda}$ changes over time, 
the potential energy also changes. We define the net work $W$ done on 
the system as the sum of the potential energy change due to the 
protocol change and the work done by the nonconservative force, i.e., 
$W(t) = \int_0^{t} dt' \dot{W}(t')$ with 
\begin{equation}
\dot{W} = \sum_{\alpha=1}^p \dot{\lambda}_\alpha (\partial_{\lambda_\alpha}
V) + \dot{\bm{x}} \circ \bm{f}.
\end{equation}
The heat dissipation is given by $Q(t)=\int_0^t dt' \dot{Q}(t')$ 
with~\cite{Sekimoto:1998uf}
\begin{equation}
\dot{Q} = \dot{\bm{x}} \circ \bm{F}.
\end{equation}
The notation $\circ$ denotes the inner product in the Stratonovich 
sense~\cite{Risken:1996vl,Gardiner:2010tp}.
The definitions of work and heat are consistent with the energy conservation law
$\dot{E} = \dot{W} - \dot{Q}$.

The total entropy production $S_{\rm tot}(t) = S_{\rm sys}(t) + S_{\rm
env}(t)$ consists of the change in 
the stochastic entropy of a system~($S_{\rm sys}(t))$ and 
the entropy production in the heat bath~($S_{\rm
env}(t)$)~\cite{Seifert:2005vb}.
The stochastic entropy is defined by $s(t) = -\ln
P(\bm{x}(t),t;\bm{\lambda}(t),\bm{\kappa}(t))$, whose mean 
value is the Shannon entropy of the probability distribution function 
$P$~\cite{Seifert:2005vb}. The change in the stochastic entropy is then
given by $S_{\rm sys}(t) = s(t)-s(0)$.
The environmental entropy production is given by the Clausius form $S_{\rm
env}(t) = Q(t)/T$~\cite{Seifert:2005vb}.

Recently, Pigolotti et al.~derived a Langevin-type stochastic differential 
equation for the total entropy 
production~\cite{Pigolotti:2017fs}.
The resulting equation reveals the universal statistical property of the
total entropy production in the nonequilibrium steady state.
Due to the steady state condition, the theory applies
only to systems driven by a time-independent force.
We extend the theory in search for the universal property of general
nonequilibrium systems even in the transient state under a time-dependent
driving force.

The total entropy production can be divided into the housekeeping 
entropy production $S_{\rm hk}$ and the excess entropy production 
$S_{\rm ex}$. 
Each contribution satisfies the fluctuation theorem 
in the absence of an odd-parity variable, such as momentum, 
under the time reversal~\cite{Esposito:2010bd,Spinney:2012uw,Lee:2013to}.
When the protocol parameters $\bm{\lambda}$ and $\bm{\kappa}$ 
are fixed to constant values, the system ultimately relaxes to  
the corresponding nonequilibrium steady state.
Maintaining the steady state, the system constantly dissipates the heat into
the heat bath and produces the entropy.  Such a contribution is called 
the housekeeping entropy production, while the rest is called the excess
entropy production. Among them, we focus on the housekeeping entropy
production. We describe below how it is defined in the single trajectory 
level. For more details, we refer readers to e.g.
Refs.~\cite{Speck:2005wp,Esposito:2010bd}.

Suppose that the position variable evolves along a trajectory 
$\bm{x}_0 \to \cdots \to \bm{x}_j \to \cdots \to \bm{x}_N$
with $\bm{x}_j = \bm{x}(t_j=j dt)$ and $dt = t/N$ while the protocols 
change as
$\bm{\lambda}_0 \to \cdots \to \bm{\lambda}_j \to \cdots \to \bm{\lambda}_N$
with $\bm{\lambda}_j = \bm{\lambda}(t_j)$ and similarly for $\bm{\kappa}$.
At each $t_j$, we can define the fictitious steady state distribution
$P_{\rm ss}(\bm{x};\bm{\lambda}_j,\bm{\kappa}_j)$ to which the system
would relax if the protocols were fixed to the values 
$\bm{\lambda}_j$ and $\bm{\kappa}_j$. 
It is given by the solution of $\partial_t P_{\rm ss} =
-\bm{\nabla}_{\bm x}\cdot {\bm J}_{\rm ss}(\bm{x};\bm{\lambda},\bm{\kappa}) = 0$
where
\begin{equation}\label{Jss}
{\bm J}_{\rm ss}(\bm{x};\bm{\lambda},\bm{\kappa}) = \mathsf{M}
\bm{F}(\bm{x},\bm{\lambda},\bm{\kappa}) P_{\rm
ss} - \mathsf{D} \bm{\nabla}_{\bm x} P_{\rm ss}
\end{equation}
is the probability current in the fictitious steady state to given
$\bm{\lambda}$ and $\bm{\kappa}$.
The housekeeping entropy production is defined as~\cite{Esposito:2010bd}
\begin{equation}\label{eq:def_Shk}
S_{\rm hk}
\simeq \sum_{j=0}^{N-1} \ln \left[\frac{P_{\rm
ss}(\bm{x}_j;\bm{\lambda}_j,\bm{\kappa}_j) \Pi(\bm{x}_j \to
\bm{x}_{j+1};\bm{\lambda}_j,\bm{\kappa}_j)}
{P_{\rm ss}(\bm{x}_{j+1};\bm{\lambda}_j,\bm{\kappa}_j) \Pi(\bm{x}_{j+1} \to
\bm{x}_{j};\bm{\lambda}_j,\bm{\kappa}_j)} \right]
\end{equation}
where $\Pi(\bm{x}\to\bm{x}';\bm{\lambda},\bm{\kappa})$ denotes the probability density of a transition from
$\bm{x}$ to $\bm{x}'$ for given instant protocols $\bm{\lambda}$ and
$\bm{\kappa}$.
The equality in \eqref{eq:def_Shk} becomes exact in the limit of
$N\to\infty$ with a fixed $t=Ndt$. The argument of the logarithm is
identically equal to unity when the transition rates satisfy the detailed
balance. Thus, the housekeeping entropy production measures the extent to
which the detailed balance is broken.
For the Langevin system described by \eqref{eq:Langevin}, the housekeeping 
entropy production is given by~\cite{Speck:2005wp}
\begin{equation}\label{eq:def_Shk2}
S_{\rm hk}(t)
= \int_0^t dt' ~ \dot{\bm{x}}(t') \circ 
\frac{\mathsf{D}^{-1} \bm{J}_{\rm
ss}(\bm{x}(t');\bm{\lambda}(t'),\bm{\kappa}(t'))}
{P_{\rm ss}(\bm{x}(t');\bm{\lambda}(t'),\bm{\kappa}(t'))}.
\end{equation}

Using the expression~\eqref{eq:def_Shk2} and the Langevin 
equation~\eqref{eq:Langevin}, 
one can obtain a Langevin-type stochastic differential equation for 
$S_{\rm hk}(t)$.
First, we rewrite the expression for the housekeeping entropy production 
in terms of the It{\^o} product instead of the Stratonovich product
to obtain
\begin{equation}
\frac{dS_{\rm hk}}{dt}
= \dot{\bm{x}}\cdot \frac{\mathsf{D}^{-1} \bm{J}_{\rm ss}}{P_{\rm ss}}
+ \bm{\nabla}_{\bm x} \cdot \left( \frac{\bm{J}_{\rm ss}}{P_{\rm ss}}
\right) .
\end{equation}
Replacing $\dot{\bm{x}}$ with the right hand side of \eqref{eq:Langevin} and
using the property $\bm{\nabla}_{\bm x} \cdot \bm{J}_{\rm ss}=0$ of
$\bm{J}_{\rm ss}$ in \eqref{Jss}, one obtains that 
\begin{equation}\label{eq:Langevin_Shk}
\frac{dS_{\rm hk}(t) }{dt}
= v(\bm{x}(t),\bm{\lambda}(t),\bm{\kappa}(t)) 
+ \sqrt{2v(\bm{x}(t),\bm{\lambda}(t),\bm{\kappa}(t))} \xi_{\rm hk}(t)
\end{equation}
where
\begin{equation}\label{v_def}
v = \frac{\bm{J}_{\rm ss} \cdot ( \mathsf{D}^{-1} 
 \bm{J}_{\rm ss})}{P_{\rm ss}^2}
\end{equation}
and $\xi_{\rm hk}(t)$ is the Gaussian white noise satisfying 
$\langle \xi_{\rm hk}(t) \rangle = 0$ and 
$\langle \xi_{\rm hk}(t) \xi_{\rm hk}(t') \rangle = \delta(t-t')$.
The noise term is equal to $\bm{J}_{\rm ss} \cdot
(\mathsf{D}^{-1} \bm{\xi}(t))/P_{\rm ss}$ with the noise $\bm{\xi}(t)$ in the
Langevin equation \eqref{eq:Langevin}. It is a Gaussian white noise with the
same statistical properties as $\sqrt{2v}\xi_{\rm hk}$.
The inner product symbol ~$\cdot$~ involving noise indicates
It{\^o} product throughout this paper.

Equation \eqref{eq:Langevin_Shk} is the key
result of this paper, from which one can discover the universal property of
the housekeeping entropy production. Before proceeding further, it is worth
comparing \eqref{eq:Langevin_Shk} with the corresponding equation for the
total entropy production considered in Ref.~\cite{Pigolotti:2017fs}.
The latter has the similar form to \eqref{eq:Langevin_Shk} with 
an additional term $(-2\partial_t \ln P)$, where $v$ is defined in
terms of the genuine probability distribution and the current 
$P$ and $\bm{J}$ instead of the fictitious steady state ones.
In the steady state, which exists only when the protocols are
time-independent constants, the additional term $(-2\partial_t \ln P)$
vanishes and the total entropy production satisfies \eqref{eq:Langevin_Shk} 
with the constant $\bm{\lambda}$ and $\bm{\kappa}$.
This comparison suggests that the housekeeping entropy production should display the 
universal statistical property even in the transient state under 
time-dependent driving forces.

\section{Universal property of the housekeeping entropy
production}\label{sec:Shk}
In this section, we explore the universal property of the housekeeping
entropy production. The universal property is guaranteed by the form of the
differential equation \eqref{eq:Langevin_Shk}. We will follow the analysis 
in Ref.~\cite{Pigolotti:2017fs} which was done for the total 
entropy production in the steady state.

The housekeeping entropy production is
known to satisfy the integral fluctuation theorem $\left\langle e^{-S_{\rm
hk}(t)}\right\rangle = 1$~\cite{Hatano:2001uc,Speck:2005wp}.
It can be derived straightforwardly from the evolution 
equation~\eqref{eq:Langevin_Shk}.
Consider a random variable $Y(t) \equiv e^{-S_{\rm hk}(t)}$ with $Y(0) =
e^{-S_{\rm hk}(0)}=1$. It satisfies
\begin{equation}\label{eq:Langevin_IFT}
\frac{dY}{dt} = -\sqrt{2 v} Y(t) \xi_{\rm hk}(t).
\end{equation}
The ensemble average of the right hand side is
identically zero due to the It\^o calculus. Therefore, one has that $\langle
Y(t)\rangle = \langle Y(0) \rangle = 1$ at any $t$, which proves the
integral fluctuation theorem.

In Eq.~\eqref{eq:Langevin_Shk}, the housekeeping entropy production has the
deterministic part $v$ and the stochastic part $\sqrt{2v} \xi_{\rm
hk}$.
Note that $v$ is deterministic in the sense that its value is fixed
for a given position variable $\bm{x}$.
Upon taking the ensemble average, the stochastic part 
$\langle \sqrt{2v} \xi_{\rm hk}\rangle$ vanishes in the It\^o calculus. 
That is, the deterministic component $v$ determines the average rate of the
housekeeping entropy production.
Note that the deterministic part \eqref{v_def} is nonnegative. 
Thus, it is useful to define a {\em housekeeping entropic time} 
$\tau$:
\begin{equation}\label{eq:def_tau_ss}
\tau(t) = \int_0^t dt' v(\bm{x}(t'),\bm{\lambda}(t'),\bm{\kappa}(t')).
\end{equation}
It is a random variable depending on the stochastic trajectory of
the system, whose mean is the average housekeeping entropy production. 
Due to the nonnegativity of $v$, the relation \eqref{eq:def_tau_ss} 
defines a one-to-one mapping between $t$ and $\tau$
for a given stochastic trajectory. Their differentials are related as
\begin{equation}\label{dtau}
d\tau = v(\bm{x}(t),\bm{\lambda}(t),\bm{\kappa}(t)) dt .
\end{equation}

We now consider the evolution of $S_{\rm hk}$ in the housekeeping
entropic time scale $\tau$. The differential form of
\eqref{eq:Langevin_Shk} is written as
\begin{equation}\label{dShk}
dS_{\rm hk} = v dt + d\mathcal{W}_{\rm hk}(t)
\end{equation}
with the uncorrelated Gaussian distributed random variables
$d\mathcal{W}_{\rm hk}(t)$
satisfying $\langle d\mathcal{W}_{\rm hk}\rangle = 0$ and $\langle
d\mathcal{W}_{\rm hk}^2\rangle = 2 v dt$.
Combining \eqref{dtau} and \eqref{dShk}, one obtains $dS_{\rm hk} = d\tau +
\sqrt{2} d\tilde{\mathcal{W}}_{\rm hk}(\tau)$ with the uncorrelated Gaussian
distributed random variables $d\tilde{\mathcal{W}}_{\rm hk}(\tau)$
satisfying $\langle
d\tilde{\mathcal{W}}_{\rm hk}\rangle = 0$ and $\langle
d\tilde{\mathcal{W}}_{\rm hk}^2\rangle = d\tau$. Equivalently, we obtain
the Langevin-type equation
\begin{equation}\label{eq:Langevin_Shk_dimless}
\frac{dS_{\rm hk}(\tau)}{d\tau} = 1 + \sqrt{2}\eta(\tau),
\end{equation}
where $\eta(\tau)$ is a Gaussian white noise satisfying
$\langle \eta(\tau) \rangle = 0$ and
$\langle \eta(\tau) \eta(\tau') \rangle = \delta(\tau-\tau')$.
Therefore, the probability distribution of the housekeeping entropy production 
for a given $\tau$ is given by the Gaussian distribution
\begin{equation}\label{eq:Gaussian}
P(S_{\rm hk}|\tau)
= \frac{1}{\sqrt{4\pi\tau}} \exp\left[ -\frac{(S_{\rm hk} - \tau)^2}{4\tau}
\right] .
\end{equation}

We emphasize that that \eqref{eq:Langevin_Shk_dimless} and
\eqref{eq:Gaussian} are valid universally for any nonequilibrium systems 
described by the overdamped Langevin equation, irrespectively of system 
details. System-specific details matter 
for the mapping between $t$ and $\tau$. However, once the housekeeping 
entropic time scale is adopted, the housekeeping entropy production always follows
the Gaussian distribution whether the system is in the steady state or a
transient state.

It is worth asking whether the excess entropy production can also be
universal after a random time transformation. 
We introduce a short-hand notation $\bm{u} \equiv \bm{J}/P$, which is
decomposed into the sum of $\bm{u}_{\rm hk} \equiv \bm{J}_{\rm ss} / 
P_{\rm ss}$ and $\bm{u}_{\rm ex} \equiv \bm{u} - \bm{u}_{\rm hk}$.
Subtracting \eqref{eq:Langevin_Shk} from the differential equation for the
total entropy production \cite{Pigolotti:2017fs}, one can derive the
differential equation for the excess entropy production $S_{\rm ex} = S_{\rm
tot} - S_{\rm hk}$. It is given by 
$$ 
\begin{aligned}
    \frac{dS_{\rm ex}}{dt}
    & = -2\partial_t \ln P - \left(
    \sum_{i=1}^p  \dot{\lambda}_i \partial_{\lambda_i} \ln P
    +\sum_{i=1}^q  \dot{\kappa}_i \partial_{\kappa_i} \ln P\right)  \\
    & ~~~ + (2\bm{u}_{\rm hk} + \bm{u}_{\rm ex}) 
    \cdot\mathsf{D}^{-1}\bm{u}_{\rm ex}
    + \sqrt{2{\bm{u}_{\rm ex}\cdot\mathsf{D}^{-1}\bm{u}_{\rm ex}}}\xi_{\rm ex}
\end{aligned}
$$
where $\xi_{\rm ex}$ is a Gaussian white noise satisfying 
$\langle \xi_{\rm ex}(t) \rangle = 0$ and 
$\langle \xi_{\rm ex}(t) \xi_{\rm ex}(t') \rangle = \delta(t-t')$.
The noise term is equal to $\bm{u}_{\rm ex}\cdot\mathsf{D}^{-1}\bm{\xi}$
with the same $\bm{\xi}$ of the Langevin equation \eqref{eq:Langevin}.
Evidently, there is no simple proportionality relation between the
deterministic part and the noise variance even when the protocol parameters 
are time-independent.
Thus, we conclude that only the housekeeping entropy production
displays the universal property in the transient state.

Meditating on the similarity between \eqref{eq:Langevin_Shk_dimless} and the
corresponding equation for $S_{\rm tot}$ in the steady 
state~\cite{Pigolotti:2017fs}, one may suspect whether the housekeeping
entropy production satisfies the thermodynamic uncertainty relation 
even in the transient regime.
Integrating \eqref{eq:Langevin_Shk}, one obtains 
$S_{\rm hk}(t) = \tau(t) + \int_0^t dt' \sqrt{2v(t')} \xi_{\rm hk}(t')$,
which yields that 
$\langle S_{\rm hk}(t)^2\rangle = \langle \tau(t)^2\rangle + 2 \langle
\tau(t) \rangle + 2 \Upsilon(t)$
with $\Upsilon(t) = \int_0^t dt' \int_0^t dt'' \langle v(t'') 
\sqrt{2 v(t')} \xi_{\rm hk}(t') \rangle$.
Thus, the Fano factor $\mathcal{F}[S_{\rm hk}(t)] =
( \langle S_{\rm hk}^2(t)\rangle - 
\langle S_{\rm hk}(t)\rangle^2 ) / \langle S_{\rm hk}(t)\rangle$
for the housekeeping entropy production is given by
\begin{equation}\label{fano}
\mathcal{F}[S_{\rm hk}(t)] = 2 + \mathcal{F}[\tau(t)] 
+ \frac{2 \Upsilon(t)}{\langle \tau(t)\rangle}  
\end{equation}
with $\mathcal{F}[\tau(t)] = (\langle \tau^2(t)\rangle - \langle
\tau(t)\rangle^2) / \langle \tau(t)\rangle$.
Following the formalism of Ref.~\cite{Pigolotti:2017fs}, one can show that
$\Upsilon(t)$ is equal to zero identically when the
system is in the steady state. Thus, one recovers the thermodynamic
uncertainty relation $(\langle S_{\rm hk}^2(t)\rangle- \langle S_{\rm
hk}(t)\rangle^2) / \langle S_{\rm hk}(t)\rangle \geq 2$ in the steady state.
In the transient state, however, \eqref{fano} does not guarantee the inequality 
$\mathcal{F}[S_{\rm hk}(t)] \geq 2$ with or without time-dependent protocol 
parameters.

\section{Numerical studies on model systems}\label{sec:Examples}
We demonstrate the universal property of the housekeeping entropy production 
with numerical simulations. The probability distributions of 
$S_{\rm hk}$ are obtained numerically in the following way: 
(i) The initial configuration $\bm{x}$ at $t=0$ is drawn from an initial 
distribution $P_{\rm ini.}(\bm{x})$.
(ii) The increments of $\bm{x}$, $\tau$, and $S_{\rm hk}$ are evaluated 
using the time-discretized equations of motion with $dt$.
The Heun algorithm is adopted~\cite{Greiner:1988tm}.
We choose $dt = 0.0001$ in simulations, which is small enough.
Time-dependent protocol parameters are also updated.
(iii) The data are collected when $t$ or $\tau$ reaches a target value.
The simulations are repeated independently for $N_{\rm samples}$ times to
construct the probability distribution.

First, we reconsider a model studied in Ref.~\cite{Pigolotti:2017fs}. 
A particle diffuses in a one-dimensional ring 
under a triangular potential
\begin{equation}
V(x) = \begin{cases}
V_0 \left(\frac{x}{x^*}\right) & {\rm for}~0 \leq x < x^*, \\
V_0 \left(\frac{1-x}{1-x^*}\right) & {\rm for}~x^* \leq x < 1 .
\end{cases}
\end{equation}
The periodic boundary condition~($x+1=x$) is applied.
In addition to the conservative force $-\partial V(x)/\partial x$, 
an $x$-independent force $f$ is applied to drive the system into a 
nonequilibrium state.
The whole system is embedded in a heat bath at temperature $T$.
The Langevin equation is given by
\begin{equation}\label{eq:Langevin_triangular}
\dot{x}(t) = \mu F(x(t)) + \sqrt{2\mu T} \xi(t)
\end{equation}
where $\mu$ is the mobility of the particle,
\begin{equation}
F(x) = \begin{cases}
F_A \equiv f - \frac{V_0}{x^*}  & {\rm for}~0 \leq x < x^*, \\
F_B \equiv f + \frac{V_0}{1-x^*} & {\rm for}~x^* \leq x < 1
\end{cases}
\end{equation}
is the total force, and $\xi(t)$ is a Gaussian white noise satisfying 
$\langle \xi(t) \rangle = 0$ and 
$\langle \xi(t) \xi(t') \rangle = \delta(t-t')$.
The protocol parameters $\bm{\lambda} = \{V_0, x^*\}$ and $\bm{\kappa} =
\{f\}$ may change over time.

The steady state probability distribution for fixed parameters is given by
\begin{equation}\label{eq:triangle_Pss}
P_{\rm ss}(x) = \begin{cases}
\alpha_1 + \alpha_2 e^{F_A x/T}
& {\rm for}~0 \leq x < x^*, \\
\alpha_3 + \alpha_4 e^{F_B x/T}
& {\rm for}~x^* \leq x < 1 ,
\end{cases}
\end{equation}
where the explicit expressions for $\alpha_i = \alpha_i(V_0,x^*,f)$ 
are found in Ref.~\cite{Pigolotti:2017fs}.
The corresponding current $J_{\rm ss} = \mu F_A \alpha_1 = \mu F_B \alpha_2$
is an $x$-independent constant. With these expressions, one is ready to
calculate the housekeeping entropy production and the housekeeping entropic
time. We choose $P_{\rm ini.}(x) = \delta(x)$ and $N_{\rm samples} = 10^6$.

\begin{figure}
\includegraphics*[width=\columnwidth]{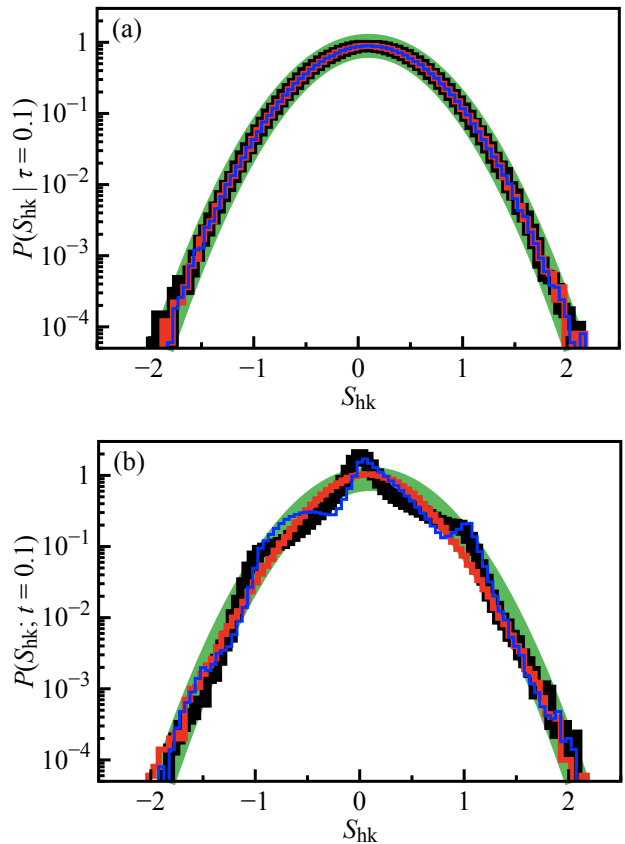}
\caption{Probability distributions of the housekeeping entropy production 
(a) at fixed $\tau=0.1$ and (b) at fixed $t=0.1$. They were obtained from 
$N_{\rm samples} = 10^6$ trajectories with $\mu=T=1$. Three different
parameter sets are considered: (i) $x^*(t) = 0.5 + 0.25 \sin(20\pi t)$,
$V_0 = 1$, and $f=1$ (black) (ii)  $f(t) = {\rm sign}(\sin(10\pi t))$,
$V_0=1$, and $x^* = 3/4$ (red) (iii)  $V_0(t) = 10t$, $x^* = 3/4$, $f=1$
(blue). Also drawn is the Gaussian distribution in \eqref{eq:Gaussian} with
$\tau=0.1$~(green).}
\label{fig1}
\end{figure}

We present the probability distribution of the housekeeping entropy
production in Fig.~\ref{fig1}. 
We performed the numerical simulations for three different protocol 
parameter sets, in which one among $V_0$, $x^*$, and $f$ varies in time 
while the others are kept to be constant.
First of all, Fig.~\ref{fig1}(a) shows the
distributions at a fixed housekeeping entropic time $\tau = 0.1$. 
Despite the difference in the protocol parameters, the probability
distributions follow the predicted Gaussian distribution 
\eqref{eq:Gaussian} with $\tau=0.1$.
For comparison, we also present the probability distributions at fixed
$t=0.1$ in Fig.~\ref{fig1}(b). The probability distributions depend on the
protocol and clearly deviate from the universal Gaussian distribution.
This example confirms the universal distribution of the housekeeping entropy
production even in the transient systems.

In order to highlight the difference of the housekeeping entropy production
and the total entropy production, 
we next consider a two-dimensional Brownian motion under a harmonic potential 
$V(\bm{x}) = \frac{1}{2}K(x_1^2 + x_2^2)$ in contact with a heat bath at
temperature $T$. A linear nonconservative force $\bm{f}(\bm{x}) = 
(\epsilon x_2, -\epsilon x_1)^{\rm T}$ drives the system into a
nonequilibrium state. The parameter $\epsilon$ stands for the strength of
the nonequilibrium driving.
For simplicity, the mobility matrix is taken to be
$\mathsf{M} = \mu \mathsf{I}$ with the identity matrix $\mathsf{I}$, and the
initial distribution of $\bm{x}$ is taken be a Gaussian with zero mean and
covariance $\sigma_0^2 \mathsf{I}$. In this model, the protocol parameters $K$ 
and $\epsilon$ are taken to be time-independent.

The model belongs to the Ornstein-Uhlenbeck process~\cite{Risken:1996vl}, 
which is exactly solvable. 
The time-dependent probability distribution is given by 
\begin{equation}
P(\bm{x},t) 
= \frac{1}{2\pi\sigma_t^2} \exp\left[ -\frac{1}{2\sigma_t^2} \left( x_1^2 + x_2^2 \right) \right]
\end{equation}
where $\sigma_t = \frac{T}{K} + e^{-2\mu K t} \left( \sigma_0^2 - \frac{T}{K} \right)$.
The corresponding probability current is
\begin{equation}
\bm{J}(\bm{x},t) = \mu P(\bm{x},t) \begin{pmatrix}
(\frac{T}{\sigma_t^2} - K)x_1 + \epsilon x_2  \\
-\epsilon x_1 + (\frac{T}{\sigma_t^2} - K) x_2
\end{pmatrix} .
\end{equation}
The steady state probability distribution is given by $P_{\rm ss}(\bm{x}) =
\frac{K}{2\pi T} \exp[-\frac{K}{2T}(x_1^2+x_2^2)]$ with the current
$\bm{J}_{\rm ss}(\bm{x}) = \mu \epsilon P_{\rm ss}(\bm{x}) (x_2, -x_1)^{\rm
T}$. Since the probability distribution in the transient state is available,
one can measure the total entropy production $S_{\rm tot}(t) = [-\ln
P(\bm{x}(t),t) + \ln P(\bm{x}(0),0)] + Q(t)/T$ as well as the housekeeping
entropy production $S_{\rm hk}(t)$ numerically.

\begin{figure}
\includegraphics*[width=\columnwidth]{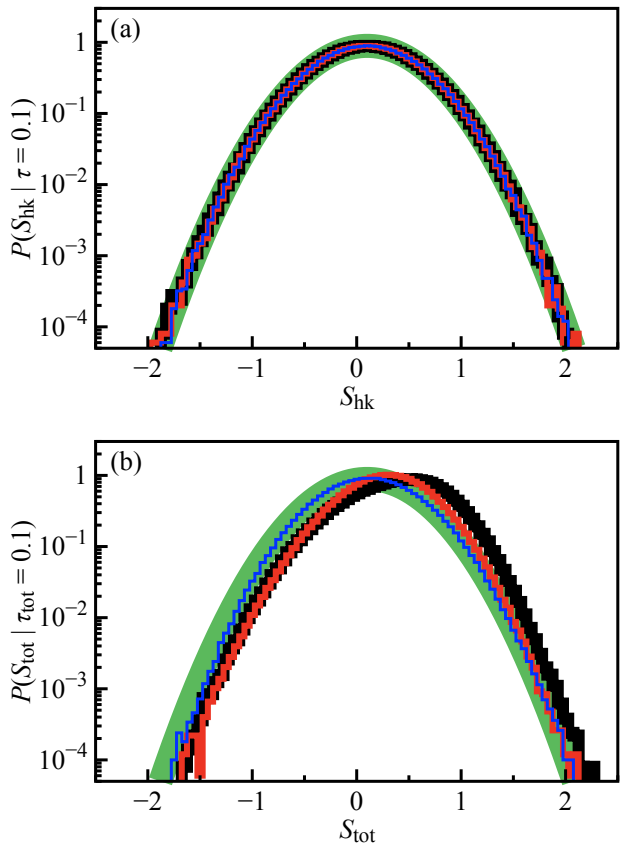}
\caption{Probability distributions of (a) $S_{\rm hk}$ at fixed 
housekeeping entropic time $\tau=0.1$ and
(b) $S_{\rm tot}$ at fixed total entropic time $\tau_{\rm tot}=0.1$.
The parameters are $\mu=T=\epsilon=1$ and $K =$ 4~(black), 6~(red), and 
8~(blue). The number of samples is $N_{\rm samples} = 10^6$.
Also shown is the Gaussian distribution in \eqref{eq:Gaussian} with
$\tau=0.1$~(green).}
\label{fig2}
\end{figure}

We present the numerical results for the total entropy production and the 
housekeeping entropy production in Fig.~\ref{fig2} with three different
values of $K = 4$, $6$, and $8$. Initially the system is prepared in
the Gaussian distribution with $\sigma_0 = 0.1 \neq \sigma_\infty \equiv  
\lim_{t\to\infty} \sigma_t$ so that the system is in a transient state. 
Then, $S_{\rm hk}$ is measured until $\tau=0.1$.
Figure~\ref{fig2}(a) shows that the housekeeping entropy production follows
the same Gaussian distribution at all values of $K$ as predicted. 
In contrast, the total entropy production is
known to be universal to a given value of the total entropic time
$\tau_{\rm tot}$ only in the steady state~\cite{Pigolotti:2017fs}.  
The total entropic time $\tau_{\rm tot}$ is defined by replacing
$P_{\rm ss}$ and $J_{\rm ss}$ of $v$ in \eqref{eq:def_tau_ss} with 
$P$ and $J$~\cite{Pigolotti:2017fs}.
We also measured the total entropy production at $\tau_{\rm tot} = 0.1$,
whose probability distribution is presented in Fig.~\ref{fig2}(b).
The probability distributions of $S_{\rm
tot}$ do not coincide with each other and do not have the Gaussian form. 
The time scale $\tau_{\rm tot}=0.1$ is too short for the system to relax into
the steady state. This example demonstrates the universal fluctuations of
the housekeeping entropy production in the transient state.

\section{Summary}\label{sec:Conclusion}

We derived a Langevin-type stochastic differential equation of the housekeeping 
entropy production for a broad class of overdamped Langevin systems.
The stochastic differential equation in \eqref{eq:Langevin_Shk} allows us to
define the housekeeping entropic time $\tau$ as given in
\eqref{eq:def_tau_ss}.
The nonnegative contribution leads to define the housekeeping entropic time.
We found that overdamped Langevin 
systems share the universal property regardless of system details:
The housekeeping entropy production follows 
the  Gaussian distribution in \eqref{eq:Gaussian} for any systems on the
entropic time scale. 
The universal property is confirmed by numerical simulations.
Our study extends the work of Ref.~\cite{Pigolotti:2017fs} significantly. 
While the total entropy production displays the universal property only in
the steady state under a time-independent protocols, the housekeeping entropy
production does even in the transient state under a time-dependent protocol.
We also remark that our formulation in \eqref{eq:Langevin_Shk} and
\eqref{eq:Langevin_Shk_dimless} provides an easy understanding of the
fluctuation theorem.

While most studies have focused on the universal property of the total entropy
production, only a few studies have considered other thermodynamic quantities.
It is noteworthy that Shiraishi et al. recently found a universal trade-off relation
between the dynamical activity of a nonequilibrium system and the excess entropy
production~\cite{Shiraishi:2018gk}.
Interestingly, similar to our results, a part of the total entropy
production is shown to uncover a universal nature of nonequilibrium 
systems.
These findings suggest that thermodynamic quantities other than the total 
entropy production can be useful in scrutinizing the universal properties 
of nonequilibrium systems. We hope that our study will promote the 
investigation of the universal properties of nonequilibrium systems not only
in the steady state but also in the transient state.

\section*{Acknowledgements}
We thank anonymous reviewers for their helpful comments to improve the
manuscript.
This work was supported by the the National Research Foundation of Korea (NRF) grant funded by
the Korea government (MSIP) (No. 2016R1A2B2013972).

\bibliographystyle{apsrev}
\bibliography{paper}

\end{document}